\documentclass[11pt,a4paper]{article}

\usepackage{physics}
\usepackage{amsmath,amssymb}
\usepackage{amsfonts}
\usepackage{jheppub}

\newcommand{\pdd}{\partial}

\newcommand{\mc}{\mathcal}

\newcommand{\mf}{\mathfrak}

\newcommand{\bbR}{\mathbb{R}}
\newcommand{\bbZ}{\mathbb{Z}}

\newcommand{\eps}{\epsilon}
\newcommand{\rp}{r_{+}}
\newcommand{\rmm}{r_{-}}

\newcommand{\qb}{\bar q}
\newcommand{\bmcL}{\bar{\mathcal L}}

\title{A generalized Selberg zeta function for flat space cosmologies}
\author{Arjun Bagchi$^1$, Cynthia Keeler$^2$, Victoria Martin$^3$ and Rahul Poddar$^4$ \\}
\affiliation[1]{Indian Institute of Technology Kanpur, Kanpur 208016, India.} 
\affiliation[2]{Department of Physics, Arizona State University, Tempe, AZ 85281, USA.}
\affiliation[3]{Department of Physics, University of North Florida, Jacksonville, FL 32224, USA.}
\affiliation[4]{Science Institute, University of Iceland, Dunhaga 3, 107, Reykjav\'ik , Iceland. \\}
\emailAdd{abagchi@iitk.ac.in, keelerc@asu.edu, victoria.martin@unf.edu, rap19@hi.is}

\abstract{Flat space cosmologies (FSCs) are time dependent solutions of three-dimensional (3D) gravity with a vanishing cosmological constant. They can be constructed from a discrete quotient of empty 3D flat spacetime and are also called shifted-boost orbifolds. Using this quotient structure, we build a new and generalized Selberg zeta function for FSCs, and show that it is directly related to the scalar 1-loop partition function. We then propose an extension of this formalism applicable to more general quotient manifolds $\mathcal{M}/\mathbb{Z}$, based on representation theory of fields propagating on this background.
  Our prescription constitutes a novel and expedient method for calculating regularized 1-loop determinants, without resorting to the heat kernel. We compute quasinormal modes in the FSC using the zeroes of a Selberg zeta function, and match them to known results.}

\begin{document}
\maketitle

\section{Introduction}
Three dimensional (3D) spacetime has been a very useful playground for understanding aspects of gravity and especially quantum gravity. The lack of propagating degrees of freedom in 3D Einstein gravity was, in the early days, thought of as an indication of the triviality of the theory. The discovery of the Ba{\~n}ados-Teitelboim-Zanelli (BTZ) black holes in 3D Anti-de Sitter spacetimes (AdS$_3$) \cite{Banados:1992wn} made it clear that 3D gravity was rich with structure but more analytically tractable than its higher dimensional avatars, making it a particularly useful tool in understanding elusive quantum aspects. 

Non-extremal BTZ black holes of generic mass and angular momentum make up the most general zero mode solutions with Brown-Henneaux \cite{Brown:1986nw} boundary conditions. BTZ black holes can be understood geometrically as orbifolds of AdS$_3$ \cite{Banados:1992gq}. For asymptotically flat spacetimes in 3D, a very similar story exists. The most generic zero mode solutions with boundary conditions outlined in \cite{Barnich:2012aw} are cosmological solutions with mass and angular momentum \cite{Cornalba:2002fi} called Flat Space Cosmologies (FSCs). FSCs can also be understood as orbifolds of flat space. Specifically the FSCs correspond to the shifted-boost orbifold of 3D flat space. The connection between the non-extremal BTZs and FSCs is actually more profound. Minkowski spacetimes can of course be reached by an infinite radius limit of AdS spacetimes. One can take a similar infinite radius limit on BTZ black holes. This rather straightforward exercise is made interesting by the fact that there are no black holes in 3D flat space. What the limit does is take the outer horizon of the non-extremal BTZ to infinity while keeping the inner horizon at a finite value \cite{Cornalba:2002fi, Bagchi:2012xr}. We thus end up with a spacetime which is the inside of the original black hole. The radial and temporal directions are flipped and hence we have a cosmological spacetime instead of a black hole. 

3D quotient manifolds (such as the BTZ black hole and FSC spacetime) are ideal for studying quantum corrections in the presence of horizons. The 3D aspect is important, because gravity is known to be 1-loop exact in this context \cite{Vassilevich:2003xt, Giombi:2008vd}. The quotient aspect is also important: in recent years, a technique has been developed to expediently calculate functional determinants of kinetic operators from the spacetime quotient structure alone \cite{Keeler:2018lza,Keeler:2019wsx,Martin:2022duk}. In this technique, one utilizes the generators of the quotient group to build a Selberg-like zeta function $Z_\Gamma(s)$, which is directly related to the regularized scalar 1-loop partition function. For example, the 1-loop partition function for a complex scalar field of mass $m$ is 
\begin{equation}\label{oneloop}
	Z^{(1)}(\Delta)=\int\mathcal{D}\phi e^{-\int\phi^*\left(-\nabla^2+m^2\right)\phi}\propto\text{det}\left(-\nabla^2+m^2\right)^{-1},
\end{equation}
where $\Delta$ is the conformal dimension of the scalar field (for AdS$_3$, we have $\Delta=1+\sqrt{1+L^2m^2}$). Traditional methods of computing \eqref{oneloop} include the heat kernel (for example, \cite{Giombi:2008vd}), which one must then regulate. Alternatively, using the Selberg technique, we directly obtain
\begin{equation}\label{zetafrac}
  Z^{(1)}_{\text{reg}}(\Delta)=\frac{1}{Z_\Gamma(\Delta)}.
\end{equation}

  The original Selberg zeta function is defined for the quotient manifold $\mathbb H^2/\Gamma$ \cite{Selberg:1956}, where $\Gamma$ is a discrete subgroup of SL$_2(\bbR)$, the isometry group of $\mathbb H^2$.
  Selberg introduced his zeta function while studying his trace formula for characters of representations of a Lie group $G$ on the space of square-integrable functions on $G/\Gamma$, where $\Gamma$ is a discrete subgroup of $G$. 
  The inputs for the function are details of the quotient, in particular the lengths and the number of primitive geodesics, for example when $\mathbb H^2/\Gamma$ is geometrically finite, the Selberg zeta function is defined as 
  \begin{equation}
    Z_{\mathbb H^2/\Gamma}(s) = \prod_{p}\prod_{n=0}^\infty (1-e^{-(s+n)\ell(p)}),
  \end{equation}
  where $p$ denotes the conjugacy classes of primitive geodesics, $\ell(p)$ is the length of the primitive geodesic $p$, and $\Gamma$ consists of the identity and hyperbolic elements only.

There is a beautiful number-theoretic aspect to the Selberg approach as well: the zeros of Selberg zeta function correspond to the quasinormal modes (QNMs) of the field in question. Thus, Selberg technique can be viewed as an extension of the work of Denef, Hartnoll and Sachdev \cite{Denef:2009kn}, who used the Weierstrass factorization theorem to cast $Z^{(1)}$ as a product over its zeros and poles. The extension is that, with the Selberg method, we seem to get the overall factor $e^{\text{Pol}(\Delta)}$ for free, without resorting to the heat kernel.

The Selberg technique was introduced in \cite{Keeler:2018lza} in the context the BTZ black hole, using the results of mathematicians Perry and Williams \cite{perry2003selberg}.
The approach was then extended to higher-dimensional hyperbolic quotients in \cite{Martin:2019flv} and for higher spin fields in \cite{Keeler:2019wsx}. An important extension to this program appeared in \cite{Martin:2022duk}, where a Selberg zeta function was built and defined for a non-hyperbolic quotient manifold (namely, the warped AdS$_3$ black hole). In addition to the physical results, the work \cite{Martin:2022duk} is of mathematical interest, potentially signalling the validity of the celebrated Selberg trace formula beyond hyperbolic quotients. Emboldened by the warped AdS$_3$ result, in this work we endeavor to construct a Selberg zeta function for flat space quotients, using the FSC as a prototypical example mirroring the BTZ solution.

In both the contexts of the BTZ black hole and the FSC spacetime, we are further able to recast the Selberg zeta function using representation theory. We conjecture a general prescription for computing  for the Selberg zeta function on general smooth quotient manifolds $\mathcal{M}/\Gamma$, where $\Gamma \sim \mathbb{Z}$:
\begin{equation}\label{IntroSelberg}
	\zeta_{\mc M/\Gamma}(s) = \prod_{\text{descendants}} \ev{1 - \gamma}_{\text{scalar primary of weight } s},
\end{equation}
where $\gamma \in \Gamma$. We are optimistic that this expression provides a tantalizing arena to further study the physical meaning of the so-called non-standard representations the appear in the calculation of Wilson spools \cite{Castro:2023dxp, Castro:2023bvo}. We will descibe this connection more fully in the Discussion section. 

The seminal analysis of Brown and Henneaux \cite{Brown:1986nw} revealed that the asymptotic symmetries of 3D asymptotically AdS spacetimes enhanced from the isometry algebra $so(2,2)$ to two copies of the Virasoro algebra 
\begin{align}
&[\mathcal{L}_n, \mathcal{L}_m ] = (n-m) \mathcal{L}_{n+m} + \frac{c}{12}n^2(n-1)\delta_{n+m,0} , \cr
&[{\bar{\mathcal{L}}}_n, {\bar{\mathcal{L}}}_m ] = (n-m) {\bar{\mathcal{L}}}_{n+m} + \frac{\bar{c}}{12} n^2(n-1)\delta_{n+m,0}, \quad [\mathcal{L}_n, {\bar{\mathcal{L}}}_m ] =0. 
\end{align}
In the above, $c$ and $\bar{c}$ are the central charges, which for Einstein gravity is given by
\begin{equation}
c = \bar{c} = \frac{3 L}{2G},
\end{equation}
where $L$ is the AdS radius and $G$ the Newton's constant. The symmetries are of course that of a 2D conformal field theory and the Brown-Henneaux analysis is often look upon as a precursor to the AdS/CFT correspondence. Due to the underlying infinite dimensional symmetries, AdS$_3$/CFT$_2$ has become a favourite testing ground for the holographic principle. BTZ black holes are dual to thermal states on the field theory side and have played (and continue to play) a starring role in lower dimensional AdS holography.  

Barnich and Compere \cite{Barnich:2006av} showed that an analysis similar to Brown and Henneaux in 3D asymptotically flat spacetime led to what is called the 3D Bondi-van der Burg-Metzner-Sachs (BMS$_3$) algebra. 
\begin{align}\label{bms3}
&[\mf L_n, \mf L_m ] = (n-m) \mf L_{n+m} + \frac{c_{\mf L}}{12}\delta_{n+m,0} \,n^2(n-1), \cr
  &[\mf L_n, \mf M_m ] = (n-m) \mf M_{n+m} + \frac{c_{\mf M}}{12}\delta_{n+m,0} \,n^2(n-1), \quad [\mf M_n, \mf M_m ] =0. 
\end{align}
The central terms can again be computed and for Einstein gravity they are given by
\begin{equation}
c_{\mf L} = 0, \quad c_{\mf M} = \frac{3}{G}.
\end{equation}
Taking a cue out of the discussion above about holography in AdS, a line of research on the construction of holography for flat spacetimes has emerged which currently is called Carrollian holography. The principle claim is that the holographic dual of 3D asymptotically flat spacetimes is a 2D quantum field theory which has the infinite dimensional BMS$_3$ algebra as its symmetries \cite{Bagchi:2010zz, Bagchi:2012cy}. The word Carrollian is reflective of the fact that the 2D field theories with this symmetry can be obtained in a Carrollian limit of 2D relativistic CFTs when one takes the speed of light to zero \cite{Duval:2014uva, Bagchi:2016bcd}. In the above context, this is the fact that the BMS$_3$ algebra can be obtained from an In{\"o}n{\"u}-Wigner contraction of the two copies of Virasoro algebra:
\begin{align}
\mf L_n = \mathcal{L}_n - {\bar{\mathcal{L}}}_{-n}, \quad \mf M_n = \frac{1}{L} \left( \mathcal{L}_n + {\bar{\mathcal{L}}}_{-n} \right), \quad L\to\infty.
\end{align}
Here the inverse of the AdS radius plays the role of the speed of light in the boundary theory and the $L\to \infty$ limit is analogous to the $c_l\to0$ limit, where $c_l$ is the speed of light. 
More generally, the BMS$_{d+1}$ algebra which one can obtain from the asymptotic analysis of $d$-dimensional flat spacetimes and the Conformal Carroll$_d$ algebra that arises in the $c_l\to0$ limit of relativistic conformal algebra in $(d-1)$- dimensions are isomorphic  \cite{Duval:2014uva, Bagchi:2016bcd}. The proposal that Carrollian CFTs are putative duals of asymptotically flat spacetimes thus passes the first and most obvious check of holography, the matching of bulk and boundary symmetries. Carrollian holography is a co-dimension one holographic dual and has recently also been used in the more physically interesting 4D asymptotically flat spacetimes, starting with \cite{Donnay:2022aba,Bagchi:2022emh}. The Carrollian approach is to be contrasted to the Celestial approach which advocates a co-dimension two holographic dual \cite{Strominger:2017zoo, Pasterski:2021raf}. A comparison between the two approaches has been recently discussed in \cite{Bagchi:2023cen}. 

This line of enquiry has led to some interesting successes in the 3D bulk -- 2D boundary case, including the matching of the bulk entropy to a BMS-Cardy analysis \cite{Bagchi:2012xr}, matching of entanglement entropy between the bulk and boundary theories \cite{Bagchi:2014iea, Jiang:2017ecm}, matching of stress-tensor correlations \cite{Bagchi:2015wna}. In a manner analogous to the BTZ black hole, the FSC solutions play a central role in the construction of holography for 3D flatspace. The entropy of the FSC is reproduced by the above mentioned BMS-Cardy analysis, which uses a Carrollian version of modular transformation \cite{Bagchi:2012xr}. One can also look at a matching of structure constants of the 2D Carrollian CFTs with an analysis of a one-point function probe in the background of a FSC. This was done in \cite{Bagchi:2020rwb} following methods similar to the AdS story \cite{Kraus:2016nwo}. The BMS-Cardy and the one-point analysis has been generalised to FSCs with extra U(1) symmetries in \cite{Bagchi:2022xug}. There is a natural generalisation to Carrollian torus two point functions and connections to bulk quasi-normal modes (QNM) in \cite{Bagchi:2023uqm}. 
 
Our investigations in this paper would address the question of QNM of the FSC solution from a completely different point of view, using the method of the Selberg zeta function which we alluded to above and go on to describe in more detail below. 
The most obvious way forward in computing say scalar QNM, i.e. by solving the Klein-Gordon equation in the FSC background, runs into problems as one now needs to put boundary conditions on the cosmological horizon as opposed to the black hole horizon. We choose to circumvent this problem by appealing to the quotient structure of the FSC and generalizing the construction of the Selberg-zeta function to these orbifolds of 3D flatspace. We will see that we will reproduce results of FSC QNM of \cite{Bagchi:2023uqm} derived earlier using Carroll modular transformation techniques. 

This paper is organized as follows. In Section 2, we review the two necessary topics for this work: the Selberg zeta function in the context of BTZ black holes and the relevant aspects of the FSC spacetime. In Section 3 we derive the Selberg zeta function for FSC spacetimes in two ways: (1) using representation theory as in equation \eqref{IntroSelberg}, and (2) from the quotient group action, as reviewed in Section \ref{subsec: RevBTZ}. In Section 4, we show that our FSC Selberg zeta function does indeed reproduce the correct scalar 1-loop partition function. In Section 5, we calculate the zeros of the FSC Selberg zeta function, and compare them to the FSC QNMs that were calculated in \cite{Bagchi:2023uqm}. In Section 6 we review our results and discuss future directions. 

\section{Review}
We review the two main ideas that we will need to construct a Selberg-like zeta function for FSC spacetimes. In Section \ref{subsec: RevBTZ}, we review how a Selberg zeta function was built for the Euclidean BTZ black hole ($\mathbb{H}^3/\mathbb{Z}$) \cite{perry2003selberg}. In Section \ref{subsec: RevFSC}, we review the geometry and quotient structure of our spacetime of interest: flat space cosmologies ($\mathbb{R}^3/\mathbb{Z}$) \cite{Cornalba:2002fi,Bagchi:2012xr,Bagchi:2013lma}.

\subsection{Selberg zeta function for $\mathbb{H}^3/\mathbb{Z}$}\label{subsec: RevBTZ}
In this section, we review how to construct the Selberg zeta function for the Euclidean BTZ black hole, which has quotient structure $\mathbb{H}^3/\mathbb{Z}$. Many of the ideas that we will discuss in this section were first presented in \cite{perry2003selberg}. 

We begin with the  BTZ black hole metric in Boyer-Lindquist-like coordinates 

\begin{equation}\label{BTZmetric}
  ds^2=-\frac{(r^2-\rp^2)(r^2-\rmm^2)}{L^2r^2}dt^2 +\frac{L^2 r^2}{(r^2-\rp^2)(r^2-\rmm^2)}dr^2 + r^2\left(d\phi-\frac{r_+r_-}{Lr^2}dt\right)^2,
\end{equation}
where $L$ is the AdS radius, and the outer and inner  horizons, $\rp$ and $\rmm$, are related to the black hole's mass $M$ and angular momentum $J$: 
\begin{equation}\label{rpm}
r_\pm = \sqrt{2GL(LM+J)} \pm \sqrt{2GL(LM-J)}.
\end{equation}

\medskip

The Euclidean BTZ black hole is obtained from \eqref{BTZmetric} via the transformations $t\rightarrow -i\tau$, $J\rightarrow -iJ_E$ and $r_-\rightarrow -i|r_-|$. The Euclidean BTZ black hole metric can be built from the Poincar\'e patch metric,
\begin{equation}\label{PoincarePatch}
	ds^2=\frac{L^2}{z^2}(dx^2+dy_E^2+dz^2),
\end{equation}
 through a set of discontinuous coordinate transformations, valid in regions $r>r_+$, $r_+>r>r_-$ and $r_->r$ \cite{Banados:1992gq}. For concreteness we will focus on the coordinate transformation valid for $r>r_+$:
\begin{equation}\label{BTZconfcoord}
  \begin{split}
    & x=\sqrt{\frac{r^2-\rp^2}{r^2-\rmm^2}}\cos\Big(\frac{\rp\tau}{L^2}+\frac{|\rmm|\phi}{L}\Big)\exp\Big(\frac{\rp\phi}{L}-\frac{|\rmm|\tau}{L^2}\Big),\\
    & y_E =\sqrt{\frac{r^2-\rp^2}{r^2-\rmm^2}}\sin\Big(\frac{\rp\tau}{L^2}+\frac{|\rmm|\phi}{L}\Big)\exp\Big(\frac{\rp\phi}{L}-\frac{|\rmm|\tau}{L^2}\Big),\\
    & z = \sqrt{\frac{\rp^2-\rmm^2}{r^2-\rmm^2}}\exp\Big(\frac{\rp\phi}{L}-\frac{|\rmm|\tau}{L^2}\Big).
  \end{split}
\end{equation}

We can now view the coordinate transformation \eqref{BTZconfcoord} through a group theoretic lens. The identification $\phi \sim \phi + 2\pi$ allows for the BTZ black hole to be understood as a quotient of AdS$_{3}$ by a discrete subgroup $\Gamma \sim \mathbb{Z}$ of the  isometry group SL$_2(\mathbb{R})\times$SL$_2,(\mathbb{R})$. We can study the group action of the single generator $\gamma\in\Gamma$ by taking $\phi \rightarrow \phi + 2\pi n$ in \eqref{BTZconfcoord}. This will map a point $(x,y_E,z) \in \mathbb{H}^3$ to another point  $(x',y_E',z')$
\begin{equation}
  \gamma^n \cdot (x,y_E,z)=(x^{\prime},y_E^{\prime},z^{\prime})
\end{equation}
through
\begin{equation}\label{BTZmap}
  \begin{split}
    x^{\prime} &= e^{2\pi n r_+ / L}\big( x \cos{(2\pi n |r_-|/L)}-y_E \sin{(2\pi n | r_-| /L)}\big),\\
    y_E^{\prime} &= e^{2\pi n r_+ /L}\big( y_E \cos{(2\pi n |r_-| /L)} + x\sin{(2\pi n |r_-|/L)} \big),\\
    z^{\prime} &= e^{2\pi n r_+ / L}z.
  \end{split}
\end{equation}
By inspecting \eqref{BTZmap}, it is clear that the group action can be understood as a dilation and a rotation:
\begin{equation} \label{eq:GroupActH3}
  \gamma
  \begin{pmatrix}
    x \\
    y_E \\
    z \\
  \end{pmatrix}
  =
  \begin{pmatrix}
    e^{2a} & 0 & 0 \\
    0 & e^{2a} & 0 \\
    0 & 0 & e^{2a} \\ 
  \end{pmatrix}
  \begin{pmatrix}
    \cos 2b_E & -\sin 2b_E & 0 \\
    \sin 2b_E & \cos 2b_E & 0 \\
    0 & 0 & 1 \\
  \end{pmatrix}
  \begin{pmatrix}
    x \\
    y_E \\
    z \\
  \end{pmatrix},
\end{equation}
where $a=\pi r_+ /L$ and $b_E=\pi |r_-| /L$. The matrix $\gamma$ has three eigenvalues: $e^{2a}$ and $e^{2a\pm 2ib_E}$.
The authors of \cite{perry2003selberg} used these eigenvalues to construct the Selberg zeta function of the BTZ black hole (which is explained in detail in section \ref{sec:zetaGpAct}):
\begin{equation}
  Z_{\Gamma}(s)=\prod_{k_{1},k_{2}=0}^{\infty}\left[1-e^{-(2a-2ib_{E})k_{1}}e^{-(2a+2ib_{E})k_{2}}e^{-2as}\right]\;\label{Eulerzeta}.
\end{equation}
In \cite{Keeler:2018lza}, it was shown explicitly that the function $Z_\Gamma(s)$ is directly related to the \textit{regularized} 1-loop partition function for a complex scalar field, obtained in a rather simple way from the spacetime quotient structure alone. As such, the authors of \cite{Keeler:2018lza} also showed that the zeros of $Z_\Gamma(s)$ are mapped to the quasinormal modes of a scalar field propagating on a BTZ background, as expected from the work of Denef, Hartnoll and Sachdev \cite{Denef:2009kn}.

We end this section with two comments. First, another useful way to write the zeta function \eqref{Eulerzeta} is
\begin{equation}
  Z_{\Gamma}(s)=\prod_{k_{1},k_{2}=0}^{\infty}\left[1-q^{k_2+s/2}\bar{q}^{k_1+s/2}\right],
\end{equation}
where $q=e^{2\pi i\tau}$, $\tau=\tau_1+i\tau_2$, $a=\pi\tau_2$ and $b_E=-\pi\tau_1$.
This has the benefit of having a symmetric contribution of $s$.
Second, it is worth mentioning that the quotient generator $\gamma\in\Gamma$ can be written as a linear combination of embedding generators \cite{Banados:1992gq}, using the conventions in appendix \ref{app:1} and \eqref{BTZmap}
\begin{equation}\label{eq:dphidef1}
  \pdd_\phi = \frac{r_+}{L} J_{U,Y} - \frac{|r_-|}{L} J_{X,V_{E}}. 
\end{equation}
Here $J_{A,B}$ are the isometries in the embedding space which preserve the AdS$_3$ hyperboloid, defined in \eqref{JAB}.
In Poincaré coordinates \eqref{PoincarePatch}, it is evident that these generate dilations $J_{U,Y}=x \partial_{x} + y_E \partial_{y_E} + z\partial_{z}$ and rotations $J_{X,V_E}=y_E \partial_{x}-x \partial_{y_E}$. 
Writing \eqref{eq:dphidef1} in terms of the SL$_2(\bbR)$ generators $\mc L_n$ and $\bar{\mc L}_n$ using \eqref{sl2rGen} and performing a Wick rotation, we have
\begin{equation}
  2\pi\pdd_\phi = 2\pi i\left(-\frac{|r_-|}{L}(\mc L_0 - \bar{\mc L_0}) + i \frac{r_+}{L} (\mc L_0 + \bar{\mc L}_0)\right). 
\end{equation}
Identifying these coefficients with $\tau_1 = -b_E/\pi = -|r_-|/L$ and $\tau_2 = a/\pi = r_+/L$, we have
\begin{equation}\label{eq:dphideftau}
  2\pi \partial_{\phi}=2\pi i\big((\mathcal{L}_0-\bar{\mc L}_0)\tau_1+i(\mathcal{L}_0+\bar{\mc L}_0)\tau_2\big).
\end{equation}
This form of writing the quotient generator is useful in generalizing the Selberg zeta function to other quotient manifolds, as seen in section \ref{repthzeta}.

\subsection{Flat Space Cosmology}\label{subsec: RevFSC}

As discussed in the introduction, flat space cosmologies can be obtained by taking a particular large $L$ limit of the BTZ black hole as well as an orbifold of 3d flat spacetimes \cite{Cornalba:2002fi,Bagchi:2012xr}. In this section, we review the limiting construction and the quotient structure of flat space cosmologies: $\mathbb{R}^3/\mathbb{Z}$ \cite{Cornalba:2002fi,Bagchi:2012xr,Bagchi:2013lma}. We will also discuss briefly properties of these solutions and their importance in the context of flatspace holography. 

\medskip

We begin by consider again the non-extremal BTZ black hole metric \eqref{BTZmetric}: 
\begin{eqnarray}
	\label{eq:btz}
	ds_{\textrm{\tiny BTZ}}^2&=&\left(8 G M-\frac{r^2}{L^2}\right) dt^2 + \frac{dr^2}{-8 G M + \frac{r^2}{  L^2} + \frac{16 G^2 J^2}{r^2}} - 8 G J dt d\phi + r^2 d\phi^2. 
\end{eqnarray}
To obtain a flat space quotient geometry from \eqref{eq:btz}, one rescales the horizons and sends the AdS radius to infinity 
\begin{equation}\label{FSCLimit}
	r_+ \to L \sqrt{8GM}  = L \hat{r}_+, \qquad r_-\to \sqrt{\frac{2G}{M}} |J| = r_0, \qquad \frac{L}{G}\to\infty,
\end{equation} 
where Newton's constant in the last expression is included to make the limit dimensionless{\footnote{It is important to note here that this rescaling of the radius with $G$ does not effect the structure of the asymptotic symmetry algebra \eqref{bms3} except changing $c_{\mf M}$ from $3/G$ to $3$.}}.  The result is the flat metric
\begin{equation}
	ds_{\textrm{\tiny FSC}}^2=\hat r_+^2 d t^2-\frac{r^2\,dr^2}{\hat r_+^2 (r^2-r_0^2)}+r^2 d\phi^2-2 \hat r_+ r_0
	dt d \phi. \label{FBTZmet}          
\end{equation}
This spacetime is known as a flat space cosmology (FSC), because the radial coordinate is now timelike, and thus the horizon $r_0$ is a timelike (i.e. cosmological) horizon.
The limit \eqref{FSCLimit} essentially pushes the outer horizon $r_+$ of the BTZ black hole to infinity, so that we are now inside the black hole, and the BTZ inner horizon $r_-$ becomes the cosmological horizon $r_0$.
The FSC metric is a solution to the vacuum Einstein equations with no cosmological constant.

Just as the BTZ black hole is a quotient of empty AdS$_3$, the FSC is a quotient of empty 3D Minkowski space. That is, for the Minkowski metric 
\begin{equation}
	ds^2=-dT^2+dX^2+dY^2,
\end{equation}
there are two coordinate transformations (either valid for $r>r_0$ or $r<r_0$) that build the FSC spacetime. The transformation valid for $r>r_0$,
\begin{equation}\label{FSCcoord}
		T=\sqrt{\frac{r^2-r_0^2}{\hat{r}_+^2}}\cosh\left(\hat{r}_+\phi\right), \quad
		X=\sqrt{\frac{r^2-r_0^2}{\hat{r}_+^2}}\sinh\left(\hat{r}_+\phi\right), \quad
		Y=r_0\phi-\hat{r}_+t,
\end{equation}
reproduces the FSC metric \eqref{FBTZmet}. Under the action $\phi\sim\phi+2\pi$, the Minkowski coordinates transform as $X^\pm\sim e^{\pm 2\pi\hat{r}_+}X^\pm$ and $Y\sim Y+2\pi r_0$, where $X^\pm\equiv X\pm T$. 

\bigskip

\noindent The Euclidean FSC solution is constructed by performing the Wick rotation \cite{Bagchi:2013lma}:
\begin{equation}
  \hat r_+ = - i \tilde r_+, \qquad t = i \tau, \qquad T = - i T_E.
\end{equation}
The quotient $\phi\to \phi + 2\pi$ is generated by the following Killing vector \cite{Bagchi:2012xr}:
\begin{equation}
  2\pi \pdd_\phi = 2\pi \big(r_0 \pdd_Y + \hat r_+ (X\pdd_T + T\pdd_X)\big) = 2\pi \big(r_0 \pdd_Y + \tilde r_+(X \pdd_{T_E}-T_E \pdd_X)\big),
\end{equation}
written in terms of the Lorentzian and Euclidean coordinates.
In terms of the BMS generators (\eqref{eq:BMSgen}, after performing a Wick rotation), the quotient generator reads
\begin{equation}
  2\pi \pdd_\phi  = 2\pi i \big(\mf L_0 \eta + i \mf M_0 \rho \big),
\end{equation}
where $\eta = \tilde r_+,~\rho = - \frac{r_0}{G}$ are the modular parameters of the flat space quotient \cite{Bagchi:2012xr}. One can also start with the quotient structure of the non-extremal BTZ and take the limit directly on the generator of the quotient to obtain the above. 

\bigskip

\noindent FSCs have many interesting properties some of which we briefly mention below. 
\begin{itemize}
\item[$\star$] The thermodynamics of FSC solutions have been studied in the literature. One can derive a first law of thermodynamics for FSC and this has peculiar negative signs \cite{Bagchi:2012xr}. The reason behind this is the cosmological horizon can be thought of as descending from the inner horizon of the non-extremal BTZ, which itself comes with a peculiar ``wrong-sign" first law \cite{Castro:2012av,Detournay:2012ug}. 

\item[$\star$] As mentioned in the introduction, one can compute the entropy of the FSC as the area of the cosmological event horizon. This can then be matched with a Cardy-like state counting computation in the boundary 2D Carrollian CFT \cite{Bagchi:2012xr} {\footnote{The initial works use Galilean CFTs instead of Carrollian CFTs. In $D=2$, the Galilean and Carrollian (conformal) algebras are isomorphic and hence computationally there is no distinction between the two. It was later appreciated that calling the algebra Carrollian is more appropriate since the isomorphism, in its most primitive version, does not hold beyond $D=2$ and the connection between Carrollian CFTs and asymptotic symmetries of flat spacetimes extend to all dimensions.}}. (See also \cite{Barnich:2012xq}.) This could be thought of as a first validation of the Carrollian holography programme. Logarithmic corrections of the FSC entropy were addressed in \cite{Bagchi:2013qva}. The BMS-Cardy formula was re-derived from the limit in \cite{Riegler:2014bia, Fareghbal:2014qga}. 

\item[$\star$] There are phase transitions between hot flat space and the FSC solutions that can be thought of as analogue of the Hawking-Page phase transitions in AdS$_3$ \cite{Bagchi:2013lma}. These cosmological phase transitions takes a time independent solution of 3d Einstein gravity to a time dependent one. 
\item[$\star$] There are generalisations of FSCs to include additional $U(1)$ charges and higher spins \cite{Gary:2014ppa, Ammon:2017vwt, Basu:2017aqn}. 
\end{itemize}

Apart from the above, FSCs have been used in other holographic contexts e.g. in studies of entanglement \cite{Bagchi:2014iea, Basu:2015evh, Jiang:2017ecm, Apolo:2020qjm} and chaos \cite{Bagchi:2021qfe} in flat spacetimes, for verifying the construction of an asymptotic formula for BMS structure constants derived from modular properties of the BMS torus one-point function \cite{Bagchi:2020rwb}, and similar studies for torus two-point functions \cite{Bagchi:2023uqm}. In general, FSCs can be thought of as the bulk duals of thermal states in the 2d Carrollian CFT. So, in conclusion, FSCs play a vital role in understanding various aspects of gravity in 3d asymptotically flat spacetimes and especially are important in understanding holography in this set-up.   

\section{Deriving the generalized Selberg zeta function for $\mathbb{R}^3/\mathbb{Z}$}

In this section, we derive a Selberg-like zeta function for the FSC spacetime in two different ways: 
\begin{enumerate}
\item a novel method using representation theory 
\item direct examination of the quotient group action, as done for the BTZ black hole in Section \ref{subsec: RevBTZ}. 
\end{enumerate}
In both cases, the Selberg-like zeta function we build is precisely related to the regularized 1-loop complex scalar partition function \cite{Barnich:2015mui} via: 
\begin{equation} \label{eq:zeta2partfn}
  Z_\Gamma(\Delta)=\frac{1}{Z^{(1)}_\text{reg}(\Delta)},
\end{equation}
where $\Delta$ labels the representation of the field. 
Thus our formalism constitutes a new and easy way to derive 1-loop partition functions of scalar fields (as well as higher spin fields \cite{Keeler:2018lza}) on quotient manifolds.

\subsection{Selberg-like zeta function from representation theory} \label{repthzeta}
We propose the following construction for a Selberg-like zeta function generalized to non-hyperbolic quotient manifolds $\mathcal{M}/\Gamma$, where $\Gamma \sim \mathbb{Z}$, motivated by representation theory\footnote{Since we consider only $\Gamma\sim\mathbb{Z}$, $\Gamma$ has only one generator. For more general subgroups of SL$_2(\mathbb{C}$), equation \eqref{GeneralSelberg} will also contain a product over generators $\gamma$.}:
\begin{equation}\label{GeneralSelberg}
  \zeta_{\mc M/\Gamma}(s) = \prod_{\text{descendants}} \ev{1 - \gamma}_{\text{scalar primary of weight } s}
\end{equation}
where $\gamma \in \Gamma$.
In the examples we consider, the quotient  $\Gamma$ is generated by the identification of the angular coordinate $\phi \sim \phi + 2\pi$, so the discrete subgroup
\begin{equation}
  \Gamma = \{\gamma^n | n\in \mathbb Z \} \subset \text{Isom}(\mc M)
\end{equation}
is generated by $\gamma = e^{2\pi \pdd_\phi}$.
Therefore, the zeta function will be expressed as 
\begin{equation}
    \zeta_{\mc M/\Gamma}(s) = \prod_{\text{descendants}} \ev{1 - e^{2\pi \pdd_\phi}}_{\text{scalar primary of weight } s}
\end{equation}
in the cases we consider. 
We will now demonstrate that this construction reproduces the Selberg zeta function for the BTZ black hole, as well as the scalar 1-loop partition function for FSCs \cite{Barnich:2015mui}.

\subsubsection{BTZ black hole}

Fields in AdS$_3$ form representations of the asymptotic symmetries of AdS$_3$, i.e. two copies of the Virasoro algebra. We will restrict our attention on the global subgroup of the algebra, which forms the AdS isometry group SL$_2(\mathbb R)\times$SL$_2(\mathbb R)$ since we consider the probe limit of the scalar field. The generators of the isometry group are $\mc L_n, \bar{\mc L}_n$, where $n = \pm1,0$ (for explicit definitions of $(\mc L_n, \bar{\mc L}_n)$, please see Appendix \ref{app:1}, and in particular \eqref{sl2rGen}).

\medskip

All states $\ket{h, \bar h}$ of the above theory are labelled by their conformal dimensions $(h,\bar{h})$: 
\begin{equation}
\mathcal{L}_{0} \ket{h, \bar h} = h \ket{h, \bar h}, \quad \bar{\mathcal{L}}_{0} \ket{h, \bar h} = {\bar h} \ket{h, \bar h}.
\end{equation}
A primary state $\ket{h, \bar h}_p$ with respect to SL$_2(\mathbb R)\times$SL$_2(\mathbb R)$ (and a ``quasi"-primary with respect to the two copies of the Virasoro algebra) is further annihilated by both $\mc L_1$ and $\bar{\mc L}_1$:  
\begin{equation}
\mathcal{L}_{1} \ket{h, \bar h}_p = 0, \quad \bar{\mathcal{L}}_{1} \ket{h, \bar h}_p = 0.
\end{equation}
We will drop the subscript $p$ on the primary state from now on. The descendants of this primary state can be written as 
\begin{equation}
	\begin{split}
		\left(\mathcal{L}_{-1}\right)^{k_2}\ket{h, \bar h}=\ket{h, \bar h, k_2}, \quad \left(\bar{\mathcal{L}}_{-1}\right)^{k_1}\ket{h, \bar h} =\ket{h, \bar h, k_1},
	\end{split}
\end{equation}
with $(k_1,k_2)\in\mathbb{Z}_{\geq 0}$. 

In terms of the SL$_2(\mathbb R)$ generators and the modular parameter for the boundary torus $\tau=\tau_1 + i \tau_2$ we have the group element $\gamma$ in terms of the generator of the quotient \eqref{eq:dphideftau}:
\begin{equation}
  \gamma = e^{2\pi \pdd_\phi} =  e^{2\pi i( (\mc L_0 - \bar{\mc L}_0 )\tau_1  + (\mc L_0+\bar{\mc L}_0)i \tau_2) } = q^{\mc L_0}\qb^{\bar{\mc L}_0},
\end{equation}
where $q = e^{2\pi i \tau}$. 
If we consider a primary scalar field of conformal dimension $\Delta=h+\bar h = 2 h = 2\bar h$ and its descendents, the eigenvalues of the SL$_2(\mathbb R)$ generators will be
\begin{equation}\label{lzeros}
	\begin{split}
		\mc L_0 \ket{h,\bar h, k_1, k_2} &= \left(h+k_2\right) \ket{h, \bar h, k_1, k_2} =\left(\frac\Delta2+k_2\right) \ket{h, \bar h, k_1, k_2} \\
		\bmcL_0 \ket{h,\bar h, k_1, k_2} &= \left(\bar h+k_1\right) \ket{h, \bar h, k_1, k_2} =\left(\frac\Delta2+k_1\right) \ket{h, \bar h, k_1, k_2}. 
	\end{split}
\end{equation}
Plugging \eqref{lzeros} into the zeta function prescription \eqref{GeneralSelberg}, we recover the Selberg zeta function for the BTZ black hole reported in \cite{perry2003selberg}, after identifying $\Delta$ with the parameter $s$:
\begin{equation}
  \begin{split}
    \zeta_{\mathbb H^3/\bbZ}(s) &= \prod_{k_1, k_2 = 0}^{\infty} \left(1 - e^{2\pi i((k_1-k_2)\tau_1+(k_1+k_2+s)i \tau_2) }\right)\\
             &=\prod_{k_1, k_2 = 0}^{\infty} \left(1 - e^{2ib_E(k_1-k_2)-2a(k_1+k_2+s) }\right). 
  \end{split}
\end{equation}
Recall that, for the BTZ black hole, the parameters $a, b$ take the values $a=\pi r_+/L,\, b_E = \pi |r_-|/L$ as discussed in Section \ref{subsec: RevBTZ}. 

\subsubsection{Flat Space Cosmology}
Now we apply the prescription \eqref{GeneralSelberg}:
\begin{equation}\label{zetagen}
	\zeta_{\mathbb R^3 /\Gamma}(s) = \prod_{\text{descendants}} \ev{1 -\gamma}_{\text{scalar primary of weight }s}
\end{equation}
to the FSC spacetime, where $\Gamma \sim \mathbb Z$.
The group element $\gamma$ in terms of the BMS$_3$ generators is
\begin{equation}
  \gamma = e^{2\pi \pdd_\phi} =  e^{2\pi i (\mf L_0\eta  + i \mf M_0 \rho) },
\end{equation}
where $\eta$ and $\rho$ are the modular parameters of the flat space quotient defined in the Section \ref{subsec: RevFSC}. 

Since fields in flat space must be representations of the global symmetry group, we can label a field with their mass $m$ and spin $k_1 - k_2$:
\begin{equation}
  \mf M_0\ket{m,k_1-k_2} = m\ket{m,k_1-k_2},\quad \mf L_0 \ket{m,k_1-k_2} = (k_1-k_2) \ket{m,k_1-k_2}. 
\end{equation}
A primary scalar field will therefore have spin $k_1 - k_2 = 0$.
We have labelled descendents with $k_1 - k_2$ because of $\mf L$'s relationship with the SL$_2(\bbR)$ $\mc L$'s \eqref{eq:VirToBMS} and the alternate construction of the zeta function discussed in the following section \ref{sec:zetaGpAct}.
Now we can plug this into \eqref{zetagen} and obtain the zeta function for FSC from just considering the quotient, once again identifying the mass $m$ with the parameter $s$:
\begin{equation}\label{FSCzetaEuc}
  \zeta_{\mathbb R^3/\mathbb Z}(s) = \prod_{k_1, k_2 = 0}^\infty\left(1-e^{2 \pi i(\eta (k_1-k_2) + i \rho s)}\right),
\end{equation}
where $\eta = \tilde r_+$ and $\rho = -r_0/G$ for FSC.

As a check, we will see that this is precisely related to the 1-loop scalar partition function on FSC \cite{Barnich:2015mui}, and will obtain it as a careful limit from the BTZ zeta function below. 

\subsection{FSC zeta function from quotient group action} \label{sec:zetaGpAct}

Another useful way to construct the zeta function is by looking at the eigenvalues of the quotient group element as an action on the coordinates (as in \eqref{eq:GroupActH3}). We begin by revisiting this technique for the BTZ black hole, including some further details regarding how this construction works in terms of prime geodesics, in Section \ref{subsubsec:3p2BTZ}. We then use this approach to build the Selberg zeta function for FSC in Section \ref{3p2FSC}.

\subsubsection{BTZ black hole}\label{subsubsec:3p2BTZ}

Let us consider the group action of $\phi \sim \phi+2\pi$ on $\mathbb H^3$, as constructed in \eqref{eq:GroupActH3}:
\begin{equation} 
  \gamma
  \begin{pmatrix}
    x \\
    y_E \\
    z \\
  \end{pmatrix}
  =
  \begin{pmatrix}
    e^{2a} & 0 & 0 \\
    0 & e^{2a} & 0 \\
    0 & 0 & e^{2a} \\ 
  \end{pmatrix}
  \begin{pmatrix}
    \cos 2b_E & -\sin 2b_E & 0 \\
    \sin 2b_E & \cos 2b_E & 0 \\
    0 & 0 & 1 \\
  \end{pmatrix}
  \begin{pmatrix}
    x \\
    y_E \\
    z \\
  \end{pmatrix}.
\end{equation}
The eigenvalues of $\gamma$ are $e^{2a}, e^{2(a\pm i b_E)}$.

It will be important to define the prime geodesic for the following discussion.
The quotient manifold $\mc M/\Gamma$ will have closed geodesics, which are geodesics that are periodic due to the action of the quotient group.
A prime geodesic is a geodesic which only requires a singular action of the group to trace out its path.
In this example, the geodesic going radially outward on the $z$ axis is the prime geodesic since this is the only geodesic which is tangent to the Killing vector generating the group action. 
The length of the prime geodesic is therefore given by the eigenvalue $e^{2a}$. 

The zeta function can be constructed by an Euler product over the eigenvalues of $\gamma$.
The weights of the eigenvalues are non-negative integers, except for the eigenvalue corresponding to the length of the prime geodesic, whose weight is the argument of the zeta function, and is not restricted to integers.
The resulting zeta function is therefore
\begin{equation}
  \zeta_{\mathbb H^3/\mathbb Z}(s) = \prod_{k_1, k_2 = 0}^{\infty}\left(1 - e^{2(a+i b) k_1}e^{2(a-i b) k_2}e^{2as}\right),
\end{equation}
which reproduces the zeta function for $\mathbb H^3/\mathbb Z$. 
This method has the advantage of not constructing representations of fields on the manifold, but the previous approach gives a direct relationship to the 1-loop partition function of the scalar field on the corresponding background.

\subsubsection{Flat Space Cosmology}\label{3p2FSC}

Similarly, let us construct the zeta function for $\mathbb R^3/\mathbb Z$.
The group action on the coordinates of flat space can be expressed as
\begin{equation}
  \gamma \begin{pmatrix}T_E\\X\\Y\end{pmatrix} =
  \begin{pmatrix}
    \cos(2\pi \eta)& \sin(2\pi \eta)  & 0\\
    -\sin(2\pi \eta)&\cos(2\pi \eta)  & 0\\
    0& 0  & e^{-2\pi \rho G \pdd_Y}\\
  \end{pmatrix}
  \begin{pmatrix}T_E\\X\\Y\end{pmatrix}. 
\end{equation}
To calculate the eigenvalue of the transformation corresponding to the translation along $Y$, it will be convenient to perform the coordinate transformation $Y = \log Z$.
The flat space metric is 
\begin{equation}
  ds^2 = dT_E^2 + dX^2 + \frac{dZ^2}{Z^2},
\end{equation}
and hence translations along $Y$ are now scale transformations along $Z$:
\begin{equation}
  Y \to Y + a \implies Z \to e^a Z. 
\end{equation}
The group action can be expressed as a matrix acting on the coordinates
\begin{equation}
  \gamma \begin{pmatrix}T_E\\X\\Z\end{pmatrix} =
  \begin{pmatrix}
    \cos(2\pi \eta)& \sin(2\pi \eta)  & 0\\
    -\sin(2\pi \eta)&\cos(2\pi \eta)  & 0\\
    0& 0  & e^{-2\pi G \rho}\\
  \end{pmatrix}
  \begin{pmatrix}T_E\\X\\Z\end{pmatrix}, 
\end{equation}
which allows us to calculate the eigenvalues of $\gamma$ as $e^{-2\pi \rho}, e^{\pm 2\pi i \eta }$.
The prime geodesic is along the $Y$ axis, which means the length of the prime geodesic is the eigenvalue $e^{-2\pi G\rho}$, and will be weighted by the argument of the zeta function.
The zeta function, following the prescription described before for the $\mathbb H^3/\mathbb Z$ case can now be constructed:
\begin{equation}
  \begin{split}
    \zeta_{\mathbb R^3/\mathbb Z}
    = \prod_{k_1, k_2 = 0}^\infty \left(1 - e^{-2\pi \rho s} e^{ 2\pi i \eta k_1}e^{-2\pi i \eta k_2}\right)= \prod_{k_1, k_2 = 0}^\infty \left(1 - e^{2\pi i (\eta(k_1 - k_2) + i \rho s)}\right), 
  \end{split}
\end{equation}
which reproduces the zeta function constructed via our representation theory method \eqref{FSCzetaEuc}.

\section{The 1-loop partition function and the Selberg zeta function}

It was shown in \cite{Keeler:2018lza}\footnote{For example, see equation (47) of \cite{Keeler:2018lza}.} that the regularized 1-loop partition function of a real, massive scalar field propagating on a BTZ black hole background is directly related to the BTZ Selberg zeta function of \cite{perry2003selberg}:
\begin{equation}\label{InverseRelation}
	Z^{\text{1-loop}}_\text{regularized}(\Delta)=\frac{1}{Z_\Gamma(\Delta)}.
\end{equation}
In the above expression, $\Delta=1+\sqrt{1+m^2L^2}$, the function $Z_\Gamma(s)$ was presented in \eqref{Eulerzeta}, and the subscript ``regularized'' is to differentiate it from the full, divergent 1-loop partition function arising from the heat kernel calculation \cite{Giombi:2008vd}:
\begin{equation}\label{DivergentEq}
	Z^{\text{1-loop}}(\Delta)=\text{Vol}\left(\mathbb{H}^3/\mathbb{Z}\right)+Z^{\text{1-loop}}_\text{regularized}(\Delta).
\end{equation} 
Equation \eqref{InverseRelation} makes it clear that the poles of $Z^{\text{1-loop}}_\text{regularized}$ correspond to the zeros of $Z_\Gamma$. 

The 1-loop partition function of a real, massive scalar field propagating on an FSC background was calculated by \cite{Barnich:2015mui}. See also \cite{Merbis:2019wgk}. In that case, equation \eqref{DivergentEq} is replaced by
\begin{equation}
	Z^{\text{1-loop}}(m)=\text{Vol}\left(\mathbb{R}^3/\mathbb{Z}\right)+Z^{\text{1-loop}}_\text{regularized}(m),
\end{equation} 
where $m$ is the scalar mass. In this section, we employ our generalized Selberg zeta function \eqref{GeneralSelberg} to obtain the interesting result 
\begin{equation}
  Z^{\text{1-loop}}_\text{regularized}(m)=\frac{1}{\zeta(m)},
\end{equation} 
in direct analogy to the BTZ black hole. Thus, equation \eqref{DivergentEq} provides a remarkably easy way to compute the regularized 1-loop partition function, eschewing heat kernel techniques.
From equation \eqref{BarnichLoop}, the 1-loop scalar partition function is 
\begin{equation}
  Z^{\text{1-loop}}_{\text{flat, scalar}}(m) =  (\det \nabla^2_{\text{flat, scalar}})^{-\frac12} = \exp(\sum_{n=1}^\infty\frac{e^{-2\pi m n \rho}}{n|(1-e^{2\pi i n \eta})|^2}).
\end{equation}
This can be rewritten to fit the form of the zeta function, 
\begin{equation}
  \begin{split}
    Z^{\text{1-loop}}_{\text{flat, scalar}}(m)
    &= \exp(\sum_{n=1}^\infty\frac{e^{-2\pi \rho mn}}{n|(1-e^{2\pi i\eta n})|^2})\\
    &= \exp(\sum_{n=1}^\infty\sum_{k_1, k_2 = 0}^\infty \frac1n (e^{-2\pi\rho m}e^{2\pi i \eta (k_1-k_2)})^n)\\
    &=\prod_{k_1, k_2 = 0}^\infty\frac{1}{1-e^{2\pi i (\eta(k_1 - k_2) + i m \rho)}} .
  \end{split}
\end{equation}
We see that the generalized Selberg zeta function constructed for $\bbR^3/\bbZ$ reproduces the regularized scalar 1-loop partition function:
\begin{equation}
  \zeta_{\bbR^3/\bbZ}(s) = (\det\nabla^2_{\text{flat, scalar}})^{\frac12} = \prod_{k_1, k_2 = 0}^\infty\left(1-e^{2\pi i( is \rho + (k_1-k_2)\eta)}\right) = \left(Z^{\text{1-loop}}_{\text{flat, scalar}}(m)\right)^{-1}\bigg|_{m=s}. 
\end{equation}


\section{The Selberg zeros as quasinormal modes}

It was shown in \cite{Keeler:2018lza} that the zeros of the BTZ Selberg zeta function give the BTZ quasinormal modes. That is, given the zeros $s_\star$ (defined by $Z_\Gamma(s_\star)=0$), the following statements are equivalent:
\begin{equation}\label{zerostoqnms}
	s_\star=\Delta \qquad \leftrightarrow \qquad \omega_{QN}=\omega_n,
\end{equation} 
where $\Delta$ is the conformal dimension of the field, $\omega_n$ are the thermal Matsubara frequencies defined by regularity at the horizon, and $\omega_{QN}$ are the QNMs. As discussed in \cite{Keeler:2018lza}, equation \eqref{zerostoqnms} constitutes an alternative method of finding quasinormal modes: given $(s_\star,\omega_n,\omega_{QN})$, we can use \eqref{zerostoqnms} to determine $\omega_{QN}$ (up to an overall function). In this section, we first calculate the thermal frequencies in FSC spacetimes, and then move on to study FSC QNMs via \eqref{zerostoqnms}.

\subsection{Thermal Frequencies in FSC}

Thermal frequencies are calculated by demanding regularity of the Euclidean metric at the horizon, and then imposing the conditions on to the scalar field solution \cite{Castro:2017mfj}.
The FSC metric is
\begin{equation}
    ds^2 = - \frac{1}{\hat r_+} \frac{r^2}{r^2-r_0^2} dr^2 + \frac{\hat r_+^2 (r^2-r_0^2)}{r^2}dt^2 + r^2 \left(d\phi -\frac{\hat r_+ r_0}{r^2} dt\right)^2.
\end{equation}
Performing the Wick rotation:
\begin{equation} \label{eq:WickRot}
  t = i \tau,\quad \hat r_+ = - i \tilde r_+,
\end{equation}
we get the Euclidean FSC metric
\begin{equation}
    ds^2 = \frac{1}{\tilde r_+} \frac{r^2}{r^2-r_0^2} dr^2 + \frac{\tilde r_+^2 (r^2-r_0^2)}{r^2}d\tau^2 + r^2 \left(d\phi -\frac{\tilde r_+ r_0}{r^2} d\tau\right)^2.
\end{equation}
We now require the FSC metric to not have any conical singularities near the horizon.
The near horizon metric can be explored using the coordinate $r^2 = r_0^2 + \epsilon \rho^2$ for small $\epsilon$, which yeilds 
\begin{equation}
    ds^2 = r_0^2 \left(d\phi - \frac{\tilde r_+}{r_0}d\tau\right)^2 + \frac{\epsilon}{\tilde r_+^2}\left(d\rho^2 + \tilde r_+^2\rho^2 d\phi^2\right) + \mc O(\epsilon^2). 
\end{equation}
For there to be no conical singularities in the subleading term, we have to go around the subleading term in $\phi$, while keeping the transverse direction $\phi - \frac{\tilde r_+}{r_0}\tau$ fixed.
Therefore,
\begin{equation}\label{eq:RegCond}
  \phi \sim \phi + \frac{2\pi}{\tilde r_+}, \quad \tau \sim \tau + \frac{2\pi r_0}{\tilde r_+^2}. 
\end{equation}
The scalar field ansatz can be written as
\begin{equation}
  \Phi(t,r,\phi) = e^{i(\omega t - k \phi)}f(r). 
\end{equation}
Performing the Wick rotation \eqref{eq:WickRot} and applying the regularity conditions at the horizon \eqref{eq:RegCond} to the scalar field, we have the condition
\begin{equation}
  e^{\frac{2\pi}{\tilde r_+^2} (i \tilde r_+ k + r_0 \omega)} = 1
\end{equation}
so that the scalar field is also regular at the horizon.
Solving the above condition for $\omega$, we obtain the thermal frequencies in the spacetime:
\begin{equation}\label{eq:thermfreq}
  \omega_n = i \frac{\tilde r_+}{r_0} (n \tilde r_+ - k), \quad n \in \bbZ. 
\end{equation}

\subsection{Quasinormal modes in FSC}
We would now like to implement the condition\footnote{Note that in flat space $\Delta=m$.}:
\begin{equation}
	s_\star=m \implies \omega_n=\omega_{QN}. 
\end{equation}
That is, tuning the Selberg zeros $s_\star$ to the mass $m$ is equivalent to equating the thermal frequencies to the quasinormal modes. We can write this condition as
\begin{equation}\label{condition}
	s_\star-m+Q\left(\omega_n-\omega_{QN}\right)=0,
\end{equation}
where $Q$ is an undetermined function. For the BTZ case, we know how to determine $Q$: we know that the BTZ QNMs should not contain the thermal integer $n$, so we can choose $Q$ to eliminate $n$.
In the FSC case it is not so simple, because (in analogy to de Sitter QNMs) we expect $\omega_{QN}$ to contain thermal and angular quantum numbers (but {\em not} radial ones).

Note that equation \eqref{FSCzetaEuc} is the Euclidean Selberg zeta for FSC.
The zeros are given by the condition 
\begin{equation}
  \begin{split}
    i s_\star \rho + (k_1-k_2)\eta = \ell
  \end{split}
\end{equation}
with $l\in\mathbb{Z}$, yielding
\begin{equation}\label{sstar}
	s_\star=\frac{\ell-\eta(k_1-k_2)}{i\rho}. 
\end{equation}
Since in the exponent of the zeta function we have the vector field generating the quotient $\pdd_\phi$, we should identify $\ell$ with the angular quantum number $k$, so that $\ell= \pm k$.
The zeros become
\begin{equation}\label{sstar2}
	s_\star=\frac{\pm k-\eta(k_1-k_2)}{i \rho}.
\end{equation}
The values of $\rho$ and $\eta$ for FSC are 
\begin{equation}\label{rhoeta}
	\rho =-\frac{r_0}{G},\quad \eta = \tilde{r}_+, 
\end{equation}
as calculated in Section \ref{subsec: RevFSC}. 
The thermal frequencies calculated by imposing regularity at the horizon are \eqref{eq:thermfreq}
\begin{equation}\label{qn1}
  \begin{split}
     \omega_n = i \frac{\tilde{r}_+}{r_0}(n \tilde{r}_+ - k).
  \end{split}
\end{equation}

Substituting equations \eqref{sstar2}, \eqref{rhoeta} and \eqref{qn1} into \eqref{condition}, we get:
\begin{equation} \label{eq:fscQNcalc}  
  \begin{split}
    \omega
    &=\omega_n-\frac{1}{Q}\left(s_\star-m\right)\\ 
    &= \frac{i \tilde{r}_+}{r_0}(n \tilde{r}_+ - k) - \frac1Q\left(-G\frac{\pm k - \tilde{r}_+ (k_1-k_2)}{i r_0} - m\right)\\
    &= \frac mQ \mp i\frac{k}{Q r_0}(G \pm Q \tilde r_+) + i \frac{\tilde r_+}{Q r_0} \big(G (k_1-k_2) + Q  n \tilde r_+\big).
  \end{split}
\end{equation}
Since we expect the FSC quasinormal modes to make sense as a limit from the BTZ quasinormal modes, we choose that the coefficient of $k$ to be zero, since the angular quantum number does not appear in the BTZ quasinormal modes.
This implies $Q = \mp G/\tilde r_+$.
Thus, the quasinormal mode reads
\begin{equation}
  \omega = \mp \tilde r_+ \left(\frac{m}{G}+ i \big((k_1-k_2)\mp n\big)\frac{\tilde r_+}{r_0} \right).
\end{equation}
Taking inspiration from the BTZ case again, we can make the identification: $k_1-k_2 = \pm n$, which recovers the leading quasinormal mode in the FSC spacetime as computed in \cite{Bagchi:2023uqm}.
\begin{equation}\label{qnm1}
  \omega = \mp \frac mG \tilde r_+ = \mp i \frac mG \hat r_+. 
\end{equation}
These correspond to the poles of a 2-point function of temporaly separated probes in the FSC boundary theory \cite{Bagchi:2023uqm}.

  We end this section with a technical note:  we have called the modes we study ``quasinormal modes'' due to their derivation from the BTZ quasinormal modes.
  However, unlike the BTZ modes, our modes are purely imaginary.
  Thus they differ from the normal modes of thermal AdS (which are purely real), as well as from the quasinormal modes of BTZ (which have both real and imaginary parts).
  In fact these modes may be more properly termed something like ``evanescent modes'' since they are purely decaying/growing, in analogy with the evanescent waves of electromagnetism.
  Regardless, since these modes still correspond to the zeroes of the zeta function, and thus to poles of the 1-loop partition function, we have chosen to stick with the terminology ``quasinormal mode'' throughout the bulk of the paper. 

\section{Discussion}

In this work, we have built a Selberg-like zeta function for flat space quotients $\mathbb{R}^3/\mathbb{Z}$, with the FSC as an interesting and concrete example.
We showed that this zeta function correctly gives the FSC scalar 1-loop partition function and the dominant QNM. These results extend the previously established Selberg method of calculating 1-loop determinants beyond negatively-curved spacetimes.
In addition, we have reinterpreted how to construct the Selberg zeta function based on representation theory of fields propagating on a given background, giving us more insight into building the Selberg zeta function in more general quotient scenarios $\mathcal{M}/\Gamma$.

There are many future directions for this work.
Most straightforwardly, it would be interesting to apply the Selberg formalism to study quantum effects in other quotient spacetimes such as k-boundary wormholes \cite{Chandra:2022fwi, Balasubramanian:2014hda, Skenderis:2009ju, Aminneborg:1997pz}, and those appearing in the context of holographic entanlgement entropy \cite{Barrella:2013wja, Faulkner:2013yia}.
It would be interesting to see the implications to corrections to black hole entropy using this formalism \cite{Banerjee:2010qc}. 

Perhaps the most intriguing quotients of all are the Lens spaces $S^3/\mathbb Z_p$ (see for example the de Sitter Farey Tail \cite{Castro:2011xb}).
This is because de Sitter admits a so-called non-standard represntation of matter, which turns out to be more closely related to quasinormal modes \cite{Castro:2020smu}.
Further, these non-standard representations are instrumental in the recently constructed Wilson spools \cite{Castro:2023dxp,Castro:2023bvo}.
Although the non-standard representations appear to be physically relevant, their precise physical description remains unknown.
It would be interesting to use \eqref{eq:zeta2partfn} to gain more insight into these non-standard representations.
This work is currently under way.

Getting back to the specific case of 3D flat spacetimes and the FSC solutions, while we have matched up the QNM answer \eqref{qnm1} that was obtained in \cite{Bagchi:2023uqm} for temporally separated probes, \cite{Bagchi:2023uqm} also obtained an answer for spatially separated modes which was more reminiscent of answers from a 2D CFT calculation. This was then curiously obtained from BTZ QNM as the sub-leading term in a $L\to\infty$ expansion. It would be of interest to see if there is a way to also obtain this curious result from our Selberg zeta function. 

\section*{Acknowledgements}

We would like to thank the Niels Bohr Institute for hospitality during the beginning of this collaboration. We thank Daniel Grumiller for comments on the manuscript. 

\medskip

AB is partially supported by a Swarnajayanti Fellowship of the Science and Engineering Research Board (SERB) and also by the following SERB grants SB/SJF/2019-20/08, CRG/2020/002035. 
CK is supported by the U.S. Department of Energy under grant number DE-SC0019470 and by the Heising-Simons Foundation “Observational Signatures of Quantum Gravity” collaboration grant 2021-2818.
VM and RP are supported by the Icelandic Research Fund under grants 195970-053 and 228952-051 and by the University of Iceland Research Fund.
VM and RP would like to thank the Isaac Newton Institute for Mathematical Sciences for support and hospitality during the programme ``Black holes: bridges between number theory and holographic quantum information'' when work on this paper was undertaken.
This work was supported by: EPSRC Grant number EP/R014604/1.

\appendix
\section*{APPENDICES}
\section{Generators and Embedding}\label{app:1}
In this appendix, we review some of the details of \cite{Cornalba:2002fi}. We describe how 3D flat space can be embedded in $\bbR^{2,2}$ as a limit of AdS$_3$. We then work out the six isometry generators $J_{A,B}$ in a series of useful coordinate systems, ultimately defining the Virasoro and BMS$_3$ generators.

We embed the hyperboloid
\begin{equation}
  -U^2+X^2+Y^2-V^2=-L^2  
\end{equation}
in  $\bbR^{2,2}$ with the metric
\begin{equation}\label{embed}
  ds^2 = -dU^2+dX^2+dY^2-dV^2. 
\end{equation}
The isometries of \eqref{embed} that preserve the hyperboloid are
\begin{equation}\label{JAB}
	 J_{A,B} = X_A \pdd_B - X_B \pdd_A,
\end{equation}
where $(A,B)$ take values $(U,V,X,Y)$.

To embed flat space as a limit of AdS$_3$ in  $\bbR^{2,2}$, we parametrize $U$ and $V$ with the compact coordinate $T/L$ by employing the following embedding:
\begin{equation}
  U = \sqrt{L^2 + X^2 + Y^2} \cos\left({\frac T L}\right), \qquad  V = \sqrt{L^2 + X^2 + Y^2} \sin\left(\frac T L\right).
\end{equation}
To obtain AdS$_3$, we should consider the universal covering space $T/L\in\mathbb{R}$, where the metric is
\begin{equation}
  ds^2 = - \frac{(L^2 + X^2 + Y^2)}{L^2} dT^2 + \frac{L^2(dX^2 + dY^2)+(XdY-YdX)^2}{L^2 + X^2 + Y^2}. 
\end{equation}
It is clear to see that in the limit $L\to \infty$ we recover three dimensional flat space:
\begin{equation}
  ds^2 = - dT^2 + dX^2 + dY^2.
\end{equation}
The rotation generators \eqref{JAB} in these ``flat'' coordinates are
\begin{equation}\label{FlatCoordGens}
  \begin{split}
    J_{U,X}&=\frac{X L}{\sqrt{L^2+X^2+Y^2}}\sin(\frac{T}{L})\pdd_T - \sqrt{L^2 + X^2 +Y^2}\cos(\frac{T}{L})\pdd_X, \\
    J_{U,Y}&=\frac{Y L}{\sqrt{L^2+X^2+Y^2}}\sin(\frac{T}{L})\pdd_T - \sqrt{L^2 + X^2 +Y^2}\cos(\frac{T}{L})\pdd_Y, \\             
    J_{X,Y}&=X\pdd_Y -Y\pdd_X,\\
    J_{U,V}&= -
             L\,\pdd_T, \\
    J_{X,V}&=\sqrt{L^2 + X^2 +Y^2}\sin(\frac{T}{L})\pdd_X+\frac{X L}{\sqrt{L^2+X^2+Y^2}}\cos(\frac{T}{L})\pdd_T, \\
    J_{Y,V}&= \sqrt{L^2 + X^2 +Y^2}\sin(\frac{T}{L})\pdd_Y+\frac{Y L}{\sqrt{L^2+X^2+Y^2}}\cos(\frac{T}{L})\pdd_T.
  \end{split}
\end{equation}

We now present the generators $J_{A,B}$ in Poincar\'e patch coordinates
\begin{equation}
	ds^2=\frac{L^2(dx^2-dy^2+dz^2)}{z^2}.
\end{equation}
The Poincar\'e coordinates in terms of the embedding coordinates are:
\begin{equation}
  x = \frac{X}{U+Y}, \quad y = \frac{-V}{U+Y},\quad z = \frac{L}{U+Y},\quad  u = y + x,\quad v = y - x.
\end{equation}
The generators \eqref{JAB} in Poincar\'e coordinates are 
\begin{equation}\label{PoinGen}
  \begin{split}
    J_{U,X} &=\frac12\left((-1+u^2-z^2) \pdd_u + (1-v^2+z^2) \pdd_v +(u-v) z \pdd_z\right),\\
    J_{U,Y} &=u\pdd_u + v\pdd_v + z\pdd_z,\\
    J_{X,Y} &=\frac12\left((-1-u^2+z^2) \pdd_u + (1+v^2-z^2) \pdd_v -(u-v) z \pdd_z\right),\\
    J_{U,V} &=\frac12\left((1+u^2+z^2) \pdd_u + (1+v^2+z^2) \pdd_v +(u+v) z \pdd_z\right),\\
    J_{X,V} &=-u\pdd_u + v\pdd_v,\\
    J_{Y,V} &=\frac12\left((-1+u^2+z^2) \pdd_u + (-1+v^2+z^2) \pdd_v +(u+v) z \pdd_z\right).              
  \end{split}
\end{equation}
If we take a particular set of linear combinations of the generators \eqref{PoinGen}, we obtain 6 generators that form two commuting copies of the SL$_2(\bbR)$ algebra. Explicitly, one can construct
\begin{equation}\label{sl2rGen}
  \begin{split}
    \mc L_{-1}&= -\pdd_u = \frac 12(J_{U,X}- J_{U,V} + J_{X,Y} + J_{Y,V}),\\
    \mc L_{0} &= -\left(u \pdd_u+\frac12 z\pdd_z\right) = \frac 12(J_{X,V} - J_{U,Y}),\\
    \mc L_{1} &= -\left(u^2\pdd_u + z^2 \pdd_v + u z \pdd_z\right) = \frac 12(-J_{U,X}- J_{U,V} + J_{X,Y} - J_{Y,V}),\\
    \bmcL_{-1}&= -\pdd_v = \frac 12(-J_{U,X} - J_{U,V} - J_{X,Y} + J_{Y,V}),\\
    \bmcL_{0} &= -\left(v \pdd_v+\frac12 z\pdd_z\right) = \frac 12(-J_{U,Y}- J_{X,V}),\\
    \bmcL_{1} &= -\left(z^2\pdd_u + v^2 \pdd_v + v z \pdd_z\right) = \frac 12(J_{U,X}- J_{U,V} - J_{X,Y} - J_{Y,V}),
  \end{split}
\end{equation}
which obey the commutation relations
\begin{equation}\label{sl2r}
  [\mc L_m, \mc L_n] = (m-n) \mc L_{m+n},\quad [\bmcL_m, \bmcL_n] = (m-n) \bmcL_{m+n}, \quad [\mc L_n, \bmcL_m] =0,  
\end{equation}
where $(m,n)$ take values $(0,\pm1)$.
For the Euclidean generators, one has to Wick rotate $V\to V_E = i V$, $\pdd_V \to \pdd_{V_E}=-i\pdd_V$ and $y\to y_E = i y$.
In the linear combinations of the embedding space generators above, this implies one must replace $J_{A,V}\to - i J_{A,V_E}$. 

The BMS$_3$ algebra (the algebra satisfied by the asymptotic symmetry group of 3D flat space) can be recovered as a limit of \eqref{sl2r} by making the following definitions
\begin{equation}\label{eq:VirToBMS}
  \mf L_{n} =  \lim_{\eps \to 0}\mc L_n - \bmcL_{-n},\quad \mf M_{n} = \lim_{\eps \to 0} \eps(\mc L_n + \bmcL_{-n}),\quad \eps = \frac{G}{L}.
\end{equation}
The BMS$_3$ generators in $T,X,Y$ coordinates read
\begin{equation}\label{eq:BMSgen}
  \begin{gathered}
    \mf L_{\pm 1}= Y(-\pdd_X \mp  \pdd_T)  + (X \mp T)\pdd_Y,\quad \mf L_{0} =( X\pdd_T + T \pdd_X) ,\\
    \mf M_{\pm 1}= G(\pdd_T \pm \pdd_X),\quad \mf M_{0} = G\pdd_Y. 
  \end{gathered}
\end{equation}
These obey the BMS$_3$ algebra without central extension
\begin{equation}
  [\mf L_m, \mf L_n] = (m-n) \mf L_{m+n},\quad   [\mf L_m, \mf M_n] = (m-n) \mf M_{m+n},\quad [\mf M_m, \mf M_n] = 0.
\end{equation}


\section{FSC partition function as a limit of the BTZ partition function}

We will see that when we carefully take this limit, the partition function poles do not accumulate into a branch cut.
This allows us to construct a meromorphic Selberg-like zeta function for FSC (for scalars, it is essentially the FSC partition function itself).
This will in turn allow us to predict FSC QNMs which are essentially guaranteed to be correct via an appropriate application of the Denef-Hartnoll-Sachdev method of computing QNMs (which is now applicable since we've established meromorphicity) \cite{Denef:2009kn}.

First, let us label the representations of the global BMS algebra as a limit of the SL$_2(\mathbb R)\times$SL$_2(\mathbb R)$ algebra, as discussed in the appendix \eqref{eq:VirToBMS},
\begin{equation}
	\mf L_n = \mc L_n - \bar{\mc L}_{-n}, \quad \mf M_n = \lim_{\frac GL\to 0}\frac{G}{L} (\mc L_n + \bar{\mc L}_{-n}). 
\end{equation}
where $\mf L$ are the diffeomorphisms of the spatial circle at null infinity, and $\mf M$ are supertranslations, and we take the dimensionless limit $\frac{G}{L}\to 0$, and $\mc L, \bar{\mc L}$ are Virasoro generators.
Fields are representations of the global symmetry algebra are labelled by the zero modes of these two sets of generators.
\begin{equation}
	\mf L_0 \ket{j,m} = j\ket{j, m}, \quad \mf M_0 \ket{j,m} = m \ket{j, m}. 
\end{equation}
We can write these eigenvalues in terms of the conformal dimensions $h,\bar h$ of the bulk field with the appropriate limit:
\begin{equation}
	j = h-\bar h, \quad m = \lim_{\frac{G}{L}\to 0} \frac{G}{L}(h + \bar h).
\end{equation}
Now let us consider the functional determinant of the scalar laplacian on AdS$_3$:
\begin{equation}
  - \log \det \nabla^2_{\text{AdS, scalar}}= 2 \sum_{n=1}^\infty \frac{q^{nh}\qb^{\,n\bar h}}{n|1-q^n|^2},
\end{equation}
where $h =\bar h$ since this is a scalar field and therefore has spin $j = 0$, $2h=\Delta$, and $q = e^{2\pi i \tau}$, $\tau = \tau_1 + i \tau_2$. 
To take the flat space limit, we have to take a limit in the modular parameter:
\begin{equation}
  \eta = \frac{\tau +\bar \tau}{2} = \tau_1,\quad  \frac{G}{L}\rho =\left(\frac{\tau - \bar \tau}{2i}\right) = \tau_2. 
\end{equation}
Now, rewriting the 1-loop scalar partition function in terms of the new modular parameter and the eigenvalues of the BMS algebra, we have
\begin{equation}
  \begin{split}
    - \log \det \nabla^2_{\text{AdS, scalar}}
    &=  2\sum_{n=1}^\infty\frac{e^{\pi i (\eta + i \eps \rho) n \frac m\eps }e^{-\pi i(\eta - i\eps\rho)n \frac m\eps}}{n(1-e^{2\pi i n (\eta + \eps \rho)})(1-e^{-2\pi i n (\eta -\eps \rho)})},\\
    &= 2\sum_{n=1}^\infty\frac{e^{-2 \pi m n\rho}}{n(1-e^{2\pi i n (\eta + \eps \rho)})(1-e^{-2\pi i n (\eta -\eps \rho)})}. 
  \end{split}
\end{equation}
where $\epsilon = \frac{G}{L}$.
We can easily take the limit $\eps \to 0$ now, to obtain
\begin{equation}\label{BarnichLoop}
	- \log \det \nabla^2_{\text{flat, scalar}}
	= 2\sum_{n=1}^\infty\frac{e^{-2\pi m n \rho}}{n|(1-e^{2\pi i n \eta})|^2}. 
\end{equation}
which is the functional determinant of the scalar Laplacian derived in \cite{Barnich:2015mui} using the heat kernel method, in which they use the notation $(\eta,\rho)\rightarrow\left(\frac{\theta}{2\pi},\frac{\beta}{2\pi}\right)$.

\bibliography{selbergfsc}
\bibliographystyle{JHEP}
\end{document}